# Observation of photonic anomalous Floquet Topological Insulators

Lukas J. Maczewsky*, Julia M. Zeuner*, Stefan Nolte, and Alexander Szameit

*Institute of Applied Physics, Abbe Center of Photonics, Friedrich-Schiller-Universität Jena, Max-Wien-Platz 1, 07743 Jena, Germany*

*These authors contributed equally to this work

**Topological insulators are a new class of materials that exhibit robust and scatter-free transport along their edges – independently of the fine details of the system and of the edge – due to topological protection [1-5]. To classify the topological character of two-dimensional systems without additional symmetries, one commonly uses Chern numbers, as their sum computed from all bands below a certain band gap is equal to the net number of chiral edge modes traversing this gap [6]. However, this is strictly valid only in settings with static Hamiltonians. The Chern numbers do not give a full characterization of the topological properties of periodically driven systems [7]. In particular, for such systems, these invariants do not determine the number of chiral edge modes within each bulk band gap. In our work, we implement a system where chiral edge modes exist although the Chern numbers of all bands are zero. We employ periodically driven photonic waveguide lattices and demonstrate topologically protected scatter-free edge transport in such anomalous Floquet topological insulators.**

The discovery of the quantized Hall effect [8] revealed the existence of a new class of extremely robust transport phenomena, which are largely independent of sample size, shape, and composition. The scatter-free nature of these phenomena can be linked to the existence of nontrivial topological invariants associated with the systems' bulk bands [2]. Shortly after the discovery of topological insulators [1,9,2,3], the concept of topology was transferred to the photonic domain of electromagnetic waves [10] with the first realization in the microwave regime implementing the photonic analogue of the quantum Hall effect [11]. The search for an optical realization of topological insulators has prompted a number of proposals [12-16], and culminated in various experimental realizations [4,5]. Photonic topological insulators may enable novel and more robust photonic devices such as waveguides, interconnects, delay lines, isolators and couplers (or anything susceptible to parasitic scattering by fabrication disorder). The field of topological photonics [6] evolved well afterwards and resulted in various further studies, such as nonlinear waves in topological insulators and the prediction of topological gap solitons [17], topological states in non-hermitian PT-symmetric media [18], topological sub-wavelength settings [19] and even systems exhibiting three-dimensional Weyl points [20].

It is commonly accepted that for two-dimensional spin-decoupled topological systems a complete topological characterization is provided by the Chern numbers of each band, which represent a set of integer topological invariants [21,22]. The number of chiral edge modes residing a band gap is given by the sum of the Chern numbers of all bands below this gap. Hence, the Chern number is equal to the difference between the chiral edge modes entering the band from below and exiting it above. However, this is strictly true only for systems that are static, that is, where the Hamiltonian is constant in time. In periodically driven (Floquet) systems, the Chern numbers employed in the static case do not give a full characterization of the topological properties. The reason is that in these systems, the fixed energy in the band structure is replaced by a periodic quasi-energy. As a consequence, the Chern numbers of all bands lying below a certain gap cannot be summed up since there exists no lowest band in the (periodic) band structure. Moreover, in such systems chiral edge modes are possible [12,23], although the Chern numbers of all bands may be zero (see Fig. 1 for an illustrative sketch). These materials are called anomalous Floquet topological insulators (A-FTI). Recently, it was shown that the appropriate topological invariants for characterizing these new phenomena are winding numbers [7], which utilize the information in the Hamiltonian for all times within a single driving period. This is in contrast to the Chern numbers of the individual bands, which only depend on the Hamiltonian evaluated over one complete driving cycle. Recently, anomalous edge states were shown in network systems that are described by a scattering matrix [24,25]. However, to date the experimental demonstration of A-FTI is still elusive.

In our work, we experimentally demonstrate an A-FTI. To this end, we work in the photonic regime and employ arrays of evanescently coupled waveguides. We consider a bipartite square lattice with two site species *A* and *B,* as it was suggested in [7]. Along the propagation direction, the structure consists of four sections with each having length *T/4*, and the entire period is T. In the first section, a particular A-site couples to neighboring B-site on its right, in the second section to its neighboring B-site above, and in the third and fourth section to its left and below, respectively (as sketched in Fig. 2a). If perfect coupling ($c_j = \frac{2\pi}{T}$) is present, this lattice structure exhibits no transport in the bulk, as an excitation is trapped by moving only in loops, whereas at the edge transport occurs (see Fig. 2b). Figure 2c shows a sketch of how we realized this lattice in our experiments. The inter-site coupling in the individual sections n is achieved by appropriately engineering directional couplers [26]. This system is described by the Hamiltonian

$$H_B(\vec{k}, z) = -\sum_{j=1}^{j=4} \begin{pmatrix} 0 & c_j(z)e^{i\vec{b}_j\vec{k}} \\ c_j(z)e^{-i\vec{b}_j\vec{k}} & 0 \end{pmatrix},$$

where the vectors $\{\vec{b}_j\}$ are given by $\vec{b}_1 = -\vec{b}_3 = (a, 0)$ and $\vec{b}_2 = -\vec{b}_4 = (0, a)$, with a being the distance between adjacent lattice sites. Additionally, for each partial step $n$, the coupling coefficients $\{c_j(z)\}$ are defined as $c_j = \delta_{jn}c$. We start our analysis by choosing the coupling coefficient $c = \frac{2\pi}{T}$, such that during each step complete coupling into the respective neighboring waveguide occurs . Obviously, the Hamiltonian is z-dependent, which for waveguide lattices is analogue to time-dependence in quantum mechanics [27] and, hence, no eigenstates exist. However, due to the periodicity in z, Floquet theory can be applied to derive a band-structure of so-called quasi-energies $\varepsilon$ [7]. Computing an effective time-independent Hamiltonian $H_{eff}$ allows a stroboscopic view of the dynamics after a full period by using $\psi(T) = U(T)\psi(0) = e^{-iTH_{eff}}\psi(0)$, where $U(T)$ is the bulk Floquet operator describing the propagation of the system after one period and $\psi(z)$ is a wave function. Consequently, the Floquet spectrum is periodic in its quasi-energies, in full correspondence to the periodicity in the transverse momentum caused by Bloch's theorem. In our lattice structure this results in two flat degenerate bands (that appear as a single band), as the bipartite character of the lattice arises only from the sequential coupling steps with four equal coupling coefficients $c_j$ and not from a sublattice potential. Since the sum of the Chern numbers of all bands has to be zero, we find that the Chern number of the flat band in our system is zero. Though, when considering a finite system, we observe the formation of chiral edge states (see Fig. 2d). In this vein, the Chern number is not the appropriate topological invariant that characterizes the existence and the amount of chiral edge states in our system. This is the very nature of an A-FTI. As it was shown earlier [7], in periodically driven systems, the topological invariant characterizing the number of chiral edge modes is the winding number $W_\varepsilon$, which is equal to the number of chiral edge modes $n_{edge}$ in a band gap at a certain quasi-energy $\varepsilon$:

$$n_{edge} = W_\varepsilon$$

The winding number is directly related to the Chern number [7]:

$$W_{\varepsilon_2} - W_{\varepsilon_1} = C_{\varepsilon_1 \varepsilon_2},$$

where $C_{\varepsilon_1 \varepsilon_2}$ is the sum of the Chern numbers of all bands residing between $\varepsilon_1$ and $\varepsilon_2$. Therefore, the difference of the number of chiral edge modes entering a band from below and exiting it above is equal to the Chern number of the respective band. We describe the approach for calculating the winding number in the Methods Section.

For our experiments, we fabricate the lattice sketched in Fig. 2c using the laser direct-writing technology [27]. For details of the fabrication, the lattice parameters, and the characterization setup we refer to the Methods Section. We start by launching light into single sites of the lattice and observe light dynamics that is summarized in Fig. 3. As clearly shown, the excited edge state travels without any scattering along edges, around corners and defects (Fig. 3a-c). This highly robust, unidirectional state is a clear signature of topological protection. However, unlike in a common Floquet topological insulator, in our system we find a flat band of bulk modes. This is shown by exciting a site in the bulk of the lattice and observing that light is trapped in a loop, indicating the excitation of only localized modes (Fig. 3d). As we observe the same dynamics for any bulk site, we can conclude that there is indeed only one band, which consists of localized degenerate states: A single flat band, which has to have a Chern number of zero. This is the unequivocal proof of having implemented an A-FTI, as clearly the Chern number does not predict the existence of the chiral edge states.

In the next step, we will analyze the impact of the inter-site coupling on the topological nature of the system. So far we considered perfect hopping ($c = \frac{2\pi}{T}$), that is, in each section $n$ the light completely couples to the neighboring site, which results in an A-FTI phase. However, when decreasing the hopping rate (which results in only partial coupling), one will eventually leave the topologically non-trivial phase [7] and enter the trivial phase exactly at $c = \frac{\pi}{T}$. This is clearly visible in the edge band structures shown in Fig. 4a,b, which shows the topological non-trivial regime in Fig.4a ($c = 1.5\frac{\pi}{T}$) with a chiral edge state being present and the trivial phase in Fig.4b ($c = 0.85\frac{\pi}{T}$). Whereas for $c > \frac{\pi}{T}$ chiral edge states exist (topological phase, Fig. 4a), at $c = \frac{\pi}{T}$ a phase transition occurs and the edge states disappear, such that for $c < \frac{\pi}{T}$ the system is in a trivial phase (Fig. 4b). Note, that in both phases the Chern number of the band is zero, and only the value of the winding number changes. In order to study this phase transition, we perform various measurements in systems with decreasing coupling constant (see Methods Section for the experimental approach). We launch light into a single site at the edge of the structure, as this populates the entire band structure, and analyze the diffraction pattern. If there is an edge state present, it is partially excited by the single site excitation and some of the evolving light will remain at the edge during propagation. However, if there is no edge state present, after a certain propagation length all of the light will have diffracted into the bulk of the system. Our experimental results are summarized in Fig. 4c, where we plot the intensity ratio $I_{edge}/I_{tot}$ as a function of the coupling constant $c$. The error bars are due to slightly fluctuating power of the writing laser and the signal to noise ratio of the recorded CCD images. For $c = \frac{2\pi}{T}$ indeed all of the light remains at the edge, as suggested by the edge band structure shown in Fig. 2d. For a decreasing coupling constant the fraction of light that remains at the edge monotonously decreases until at $c < \frac{\pi}{T}$ no edge states is present, as the trivial phase is reached.

Importantly, the region where chiral edge states are found reduces for decreasing coupling strength: Whereas in the center of the edge band structure around $k_x = 0$ such states are always found until the phase transition occurs, modes close to the edge of the band structure (around $k_x = \frac{\pi}{2a}$) continuously cease to exist for decreasing coupling. This is illustrated when exciting the band structure only at the specific momentum $k_x = \frac{\pi}{2a}$ by using an appropriately tilted broad beam [28]. In Figs. 5a-c, it is shown that the state completely remains at the edge of the lattice for $c = 2\frac{\pi}{T}$ (Fig. 5a), and partially spreads into the bulk for $c = 1.63\frac{\pi}{T}$ while a significant fraction is still trapped at the edge (Fig. 5b). However, for $c = 1.2\frac{\pi}{T}$ the light almost completely diffracts away from the edge as no chiral edge modes remain at $k_x = \frac{\pi}{2a}$ for this low coupling strength (Fig. 5c). Note that the chiral edge states reside on every second waveguide solely, such that we excited only those with the broad beam. Our results are summarized in Fig. 5d, where the fraction of the light trapped at the lattice edge is plotted as a function of the coupling strength. One clearly sees the drop in light intensity at the edge, proving the disappearance of the edge states for decreasing coupling strength.

Our experimental observation of an A-FTI opens a new chapter in the field of topological physics. The results presented here clearly demonstrate the significance of the winding number as the appropriate topological invariant characterizing periodically driven systems. Moreover, the chiral edge states in A-FTI are highly robust to distortions in the lattice structure (including defects and imperfect hopping). But what is the impact of nonlinearity on the formation of these chiral edge states? Does the dimensionality play a significant role? What are the possibilities to obtain different phases than reported here? The answer to these and other intriguing questions are now in reach.

**Methods:**

**Winding number**

To calculate the number of chiral edge modes in a periodically driven system, the behavior of the system during a full driving period has to be taken into account, by employing the time evolution operator $U(\vec{k},t) = T\exp\left(-i\int_0^t dt' H(\vec{k},t')\right)$, with $\vec{k}$ as the momentum. In a system exhibiting a flat band at quasi-energy $\varepsilon = 0$ the winding number $W$ can be calculated as [7]:

$$n_{edge} = W[U] = \frac{1}{8\pi^2}\int dt\, dk_x\, dk_y \cdot Tr\left(U^{-1}\partial_t U \cdot \left[U^{-1}\partial_{k_x}U, U^{-1}\partial_{k_y}U\right]\right).$$

In the case of curved (dispersive) bands, the winding number in a gap is $W_\varepsilon = W[U_\varepsilon]$, with $U_\varepsilon$ being constructed as follows [7]:

$$U_\varepsilon(\vec{k},t) = \begin{cases} U(\vec{k},2t) & \text{if } 0 \leq t \leq \frac{T}{2} \\ V_\varepsilon(\vec{k},2T-2t) & \text{if } \frac{T}{2} \leq t \leq T, \end{cases}$$

Here, $V_\varepsilon(\vec{k},t) = \exp(-iH_{eff}(\vec{k})t)$ with $H_{eff}(\vec{k}) = \frac{i}{T}\log U(\vec{k},T)$. The branch cut of the logarithm is chosen such that:

$$\log e^{-i\varepsilon T + i0^+} = -i\varepsilon T$$
$$\log e^{-i\varepsilon T + i0^-} = -i\varepsilon T - 2\pi i.$$

**Fabrication of the structures**

The waveguides were written [27] inside a high-purity 15cm long fused silica wafer (Corning 7980) using a RegA 9000 seeded by a Mira Ti:Al$_2$O$_3$ femtosecond laser. Pulses centered at 800nm with duration of 150fs were used at a repetition rate of 100kHz and energy of 450nJ. The pulses were focused 671µm to 883µm under the sample surface using an objective with a numerical aperture (NA) of 0.35 while the sample was translated at constant speed of 100mm/min by high-precision positioning stages (ALS130, Aerotech Inc.). The mode field diameters of the guided mode were 10.4μm × 8.0μm at 633nm. Propagation losses and birefringence were estimated at 0.2dB/cm and in the order of $10^{-7}$, respectively. The waveguides are equally spaced by 40μm to ensure that there is no unwanted coupling between them. In the individual sections of the lattice (shown in Fig. 2c) the waveguides that couple converge to a spacing of 9.7μm to ensure significant inter-site hopping. For obtaining different coupling strengths the length of the coupling region was appropriately designed [26].

**Characterization of the structures**

For the observation of the light evolution, light from a tunable Helium Neon laser (Thorlabs HTPS-EC-1) was launched into the system using a NA=0.35 objective. Whereas this is sufficient for single-site excitation, for the broad excitation the beam was expanded with a slit and a biconvex lens (f=35mm) perpendicular to the orientation of the slit. Together with every other waveguide starting 2cm later in propagation direction and an appropriate tilt of the sample we excite the correct transverse momentum [28]. We fabricated several structures with different coupling strengths as described above. However, in order to achieve more data points, we used different excitation wavelengths (633nm, 604nm, 594nm, 543nm) that allowed us to further manipulate the coupling strength [29]. We calibrated the wavelength-dependent coupling strength for the different interaction lengths of the individual sections in independent directional couplers.

**Acknowledgements:**

The authors gratefully acknowledge financial support from the Deutsche Forschungsgemeinschaft (grants SZ 276/7-1, SZ 276/9-1, BL 574/13-1, GRK 2101/1) and the German Ministry for Science and Education (grant 03Z1HN31).

**Author contributions:**

L. M. performed the measurements, J. Z. elaborated on the theory, A. S. supervised the project. All authors discussed the results and co-wrote the paper.

**Competing financial interests:**

There are no competing financial interests.

**Figure Captions:**

**Fig. 1:** Conceptual sketch of the band structure in a driven system, which is periodic in momentum $k$ and quasi-energy $\varepsilon$. Essentially, the band structure is analog to a torus (see inset). This allows chiral edge modes to exist even if the Chern numbers of all bands are equal to zero.

**Fig. 2:** Bipartite lattice structure with periodic driving. (a) The coupling to the neighboring waveguides occurs in four steps of equal length; in each step, hopping takes place solely along the highlighted bonds with a coupling strength $c_j$; all other couplings are zero. (b) If the coupling during each step is 100% ($c_j = \frac{2\pi}{T}$), after a full driving period T, one observes the formation of localized bulk modes without dispersion and chiral edge modes travelling along the lattice boundaries. (c) A schematic sketch of the fabricated sample, in which the waveguides are drawn pairwise together to enable evanescent coupling. The initial waveguide spacing is $a = 40$ μ$m$ ensuring negligible coupling between adjacent guides. (d) The edge band structure of periodic quasi-energies in the case $c_j = \frac{2\pi}{L}$, exhibiting a flat bulk band and dispersionless chiral edge modes.

**Fig. 3:** Light evolution in the lattice after single-waveguide excitation. Evolution of the excited chiral edge state (a) along the edge, (b) around a corner, and (c) along artificial defects in the lattice structure. (d) If a bulk waveguide is excited, light follows a loop

trajectory, as only localized flat-band modes are excited. The excited guide is marked with red, the output guide with white, and the trajectory is visualized with a white arrow.

**Fig. 4:** Phase transition between the trivial and the topological phase. (a) If the coupling coefficients are reduced to $c = 1.5\frac{\pi}{T}$, the bulk band is not flat anymore. However, chiral edge modes still exist in the center of the Brillouin zone, and the winding number remains $W_{\pi/T} = 1$. (b) For $c = 0.85\frac{\pi}{T}$, the chiral edge states have disappeared, and the winding number is $W_{\pi/T} = 0$. (c) Visualization of the phase transition using a single waveguide excitation. As this populates all modes, the fraction of the trapped light at the surface as a function of the coupling strength indicates the amount of existing chiral edge modes. When the coupling is $c < \frac{\pi}{T}$, these modes disappear and essentially all light diffracts into the bulk. The different topological phases are marked with orange ($W_{\pi/T} = 1$) and red ($W_{\pi/T} = 0$), in both regimes, however, the Chern number is zero.

**Fig. 5:** The region where chiral edge modes exist in the band structure reduces with decreasing coupling strength $c$. This is shown by exciting the band structure at $k_x = \frac{\pi}{2a}$ using an appropriately tilted broad beam and observing the diffraction pattern for (a) $c = \frac{2\pi}{T}$, (b) $c = 1.63\frac{\pi}{T}$, and (c) $c = 1.2\frac{\pi}{T}$. The excited waveguides are marked in red, the integration area in white. (d) The plot of the fraction of the trapped light at the edge clearly indicates that at $k_x = \frac{\pi}{2a}$ the amount of intensity exciting edge states significantly reduces for decreasing coupling strength.

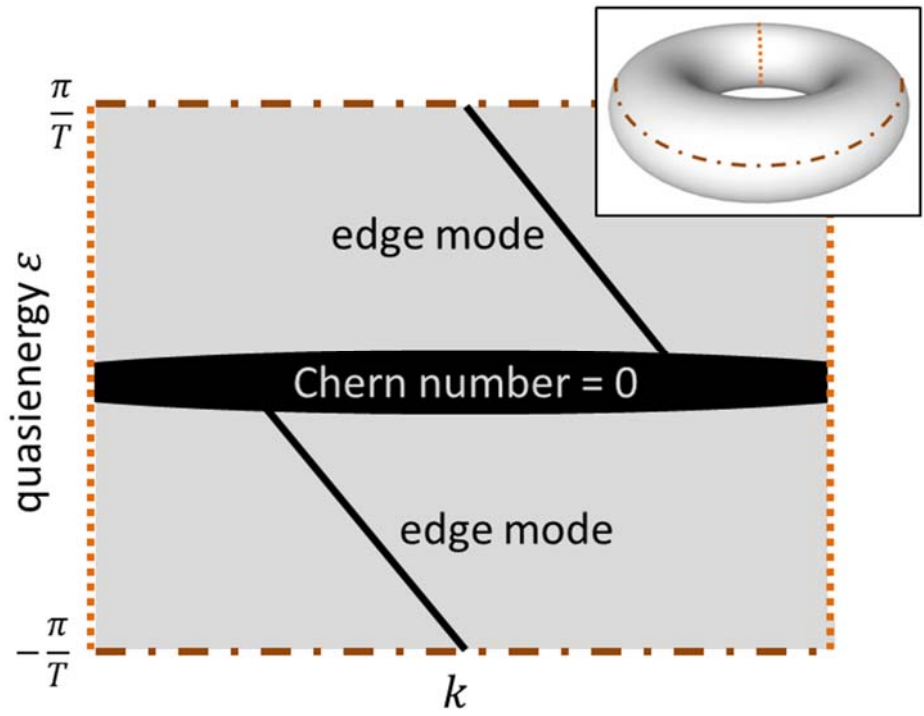

Figure 1

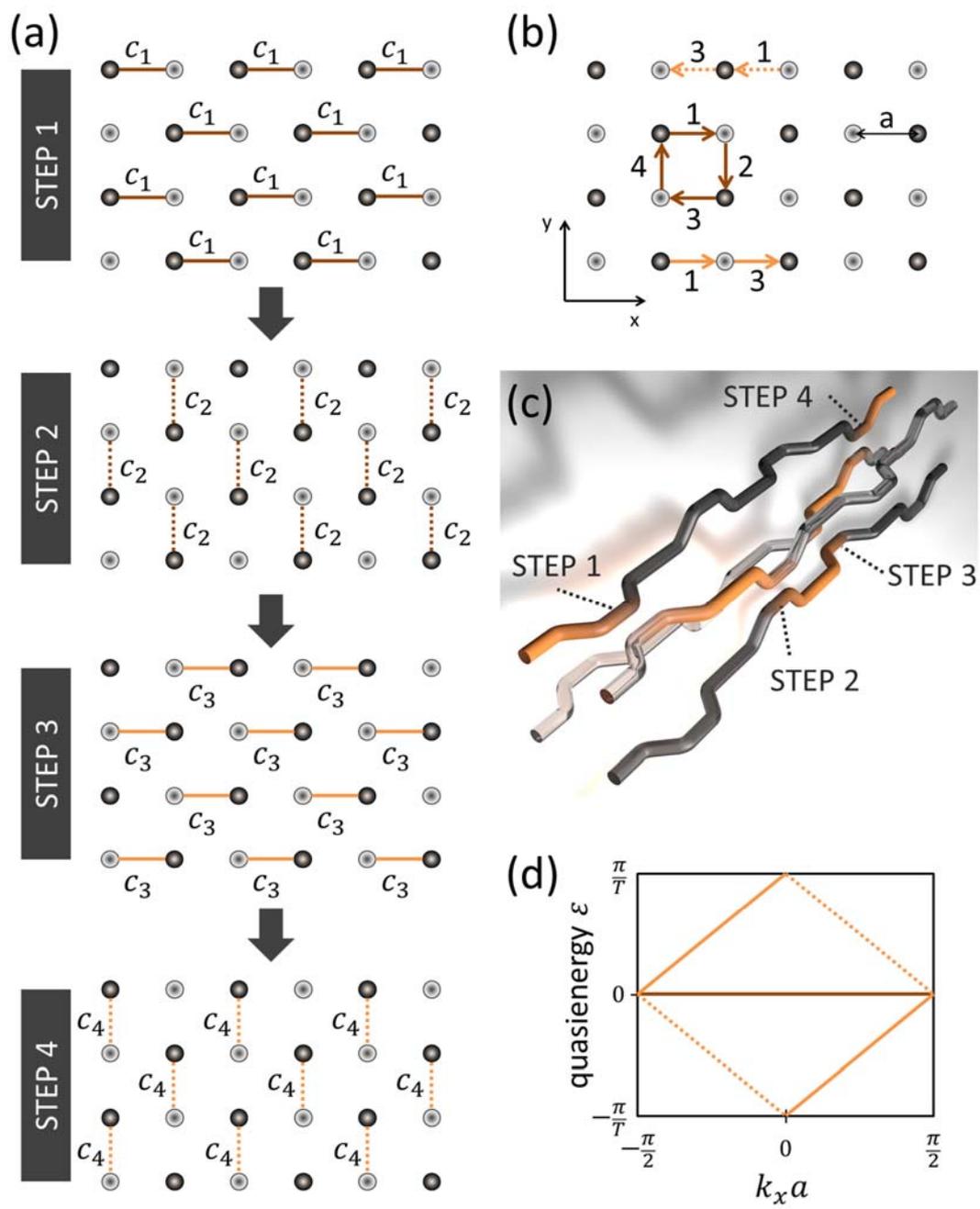

Figure 2

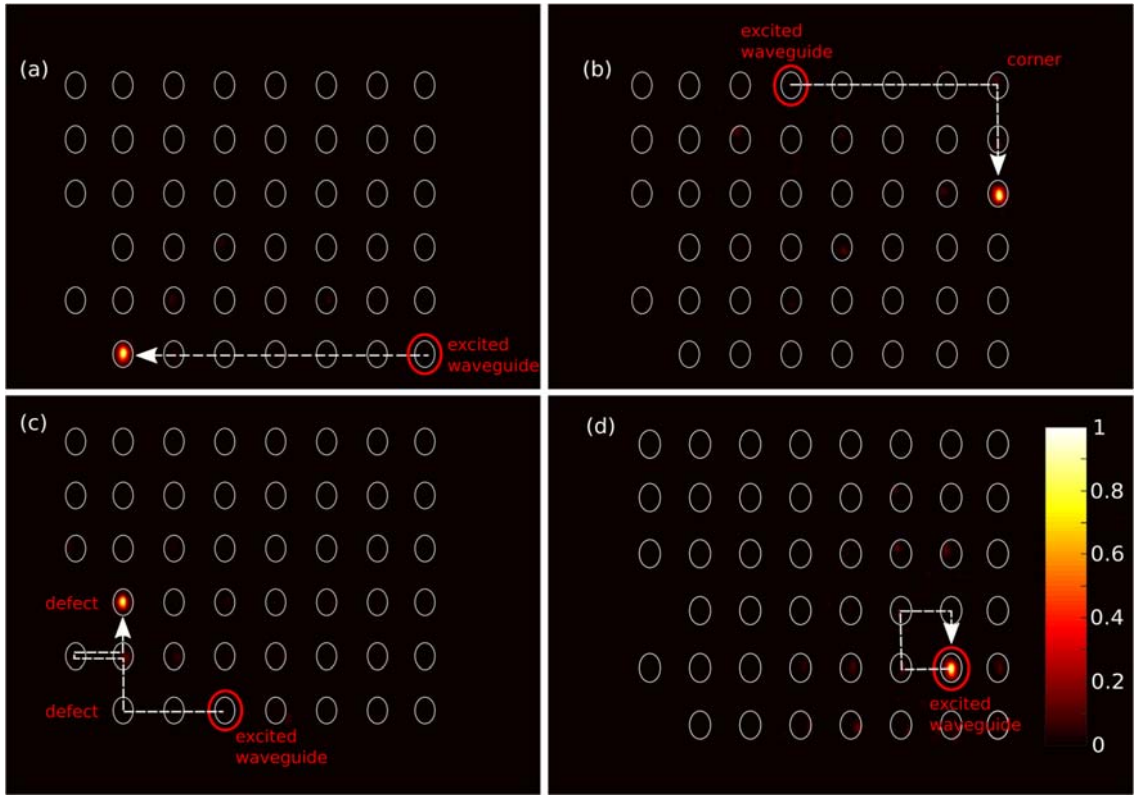

Figure 3

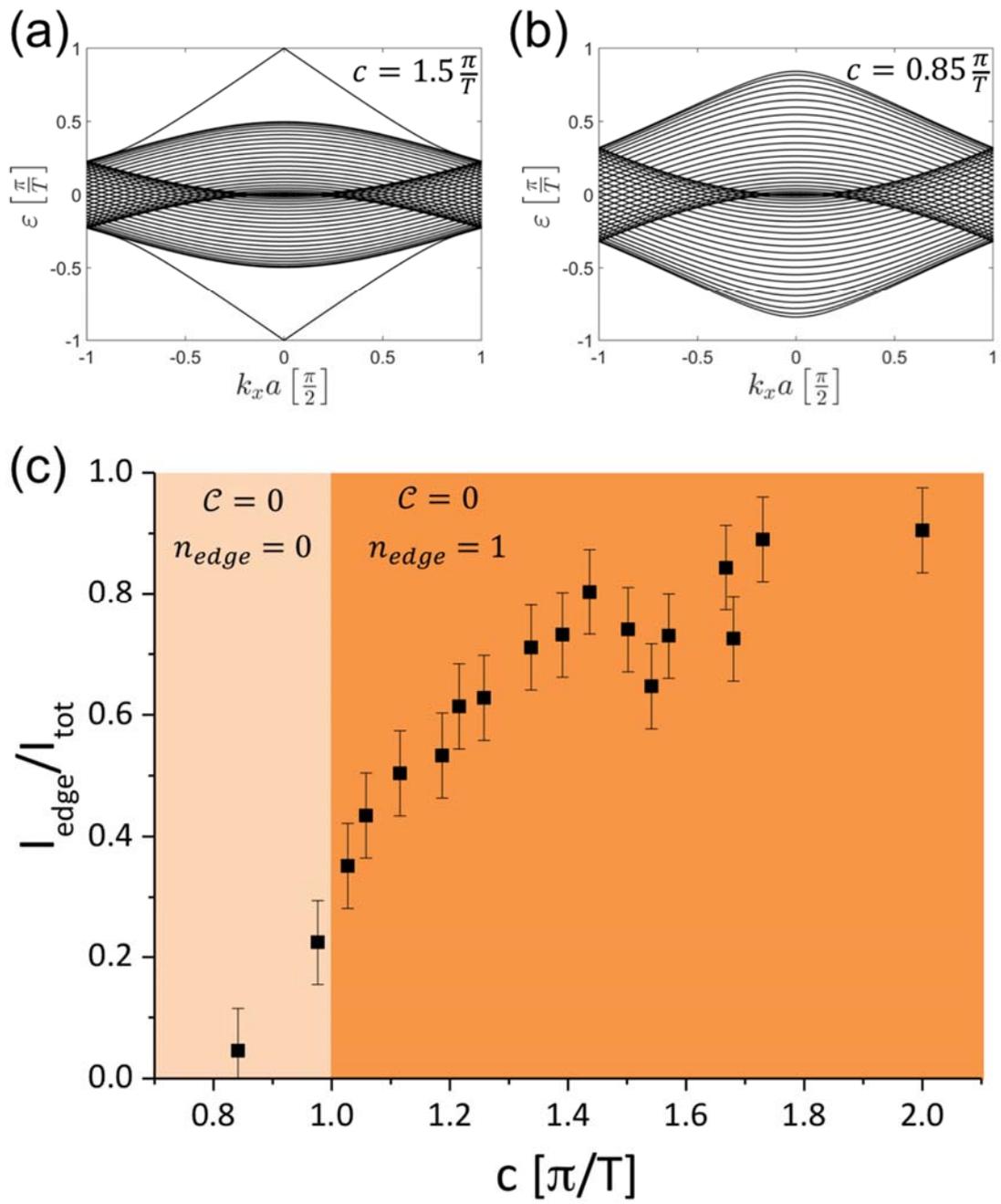

Figure 4

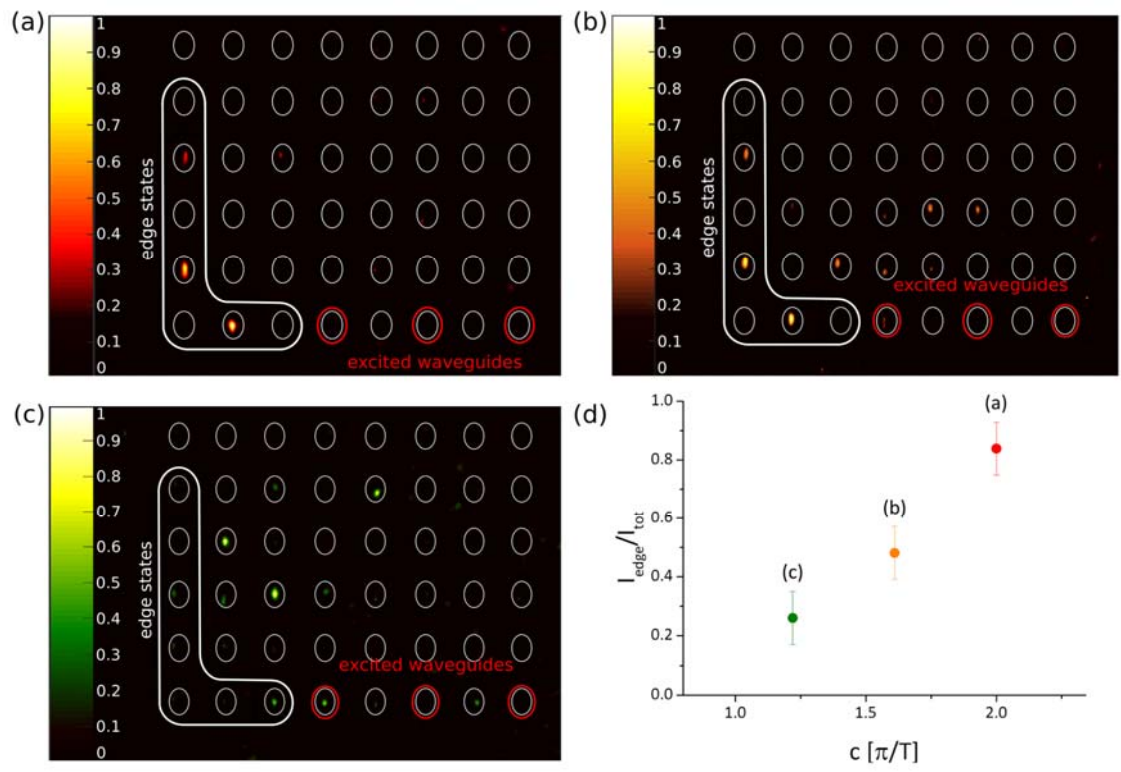

Figure 5